\documentclass[notitlepage,12pt,aps,pre,showpacs,superscriptaddress,groupedaddress,amsmath,amssymb]{revtex4-1}
\usepackage{epsfig}
\usepackage{graphics}
\usepackage{lipsum}
\usepackage{ulem}
\usepackage{sidecap}
\usepackage{graphicx} %Include figure files
\usepackage{dcolumn} %Align table columns on decimal point
\usepackage{bm}% bold math
\usepackage{longtable}
\usepackage{xcolor,colortbl}
\usepackage{color, colortbl}

\begin{document}
\title{Analysis of COVID-19 in India using  vaccine epidemic model incorporating vaccine effectiveness and herd immunity}
\author{V.~R.~Saiprasad$^{1}$,  R. Gopal$^{1}$, V. K. Chandrasekar$^{1}$ and M. Lakshmanan$^{2}$}
\address{$^1$Department of Physics, Centre for Nonlinear Science \& Engineering, School of Electrical \& Electronics Engineering, SASTRA Deemed University, Thanjavur -613 401, Tamilnadu, India.\\
$^2$ Department of Nonlinear Dynamics, School of Physics, Bharathidasan University, Tiruchirappalli -620 014, Tamil Nadu, India.\\}

\date{\today}
\begin{abstract}
\par COVID-19 will be a continuous threat to human population despite having a few vaccines at hand until we reach the endemic state through natural herd immunity and total immunization through universal vaccination. However, the vaccine acts as a practical tool for reducing the massive public health problem and the emerging economic consequences that the continuing COVID -19 epidemic is causing worldwide, while the vaccine efficacy wanes. In this work, we propose and analyze an epidemic model of  Susceptible-Exposed-Infected-Recovered-Vaccinated (SEIRV) population taking into account the rate of vaccination and vaccine waning.  The dynamics of the model has been investigated, and the condition for a disease-free endemic equilibrium state is obtained. Further, the analysis is extended to study the COVID-19 spread in India by considering the availability of vaccines and the related critical parameters such as vaccination rate,  vaccine efficacy and waning of vaccine’s impact on deciding the emerging fate of this epidemic. We have also discussed the conditions for herd immunity due to vaccinated individuals among the people.  Our results highlight the importance of vaccines, the effectiveness of booster vaccination in protecting people from infection, and their importance in epidemic and pandemic modelling.
\end{abstract}

%\pacs{COVID-19 and SEIRV model and  Vaccine waning and herd immunity}

\maketitle

\section{Introduction}
Mathematical models in epidemiology help in estimating the spread of infection and identifying the possible outcomes of an epidemic\cite{org,matt,he}. Since the onset of SARS-Cov2 (COVID-19), this kind of mathematical models have become essential in studying and predicting the disease dynamics which also act as  reliable tools for healthcare systems to make critical decisions. In particular, populations are divided into segments based on their health condition in compartmental epidemiological models. The SEIR model, for example, divides the population into four sub-population compartments: Susceptible, Exposed, Infectious, and Recovered (SEIR). These models are used to forecast epidemiological metrics including disease transmission, total number of infections, and epidemiological curve form. In recent times, the SEIR model has been extended with multiple compartments to predict the nature of  COVID-19 in the literature~\cite{he,lin,li,gopal,gopal2,post,kavi,chen,ranjan,savi,nld}.

More recently, the SEIR model~\cite{he,lin,li} was used to examine the number of infected individuals in India during the initial lock-down and unlocked periods, starting from March 25, 2020, up to October 31, 2020. Further, the analysis has also been extended by three of the present authors to November 21, 2021, with the help of the initial transmission rate of COVID-19 by considering the initial number of infected people in the country~\cite{gopal,gopal2,gov1}. The predictions of the number of infected individuals from our studies  agreed with the actual data of the daily rate of the number of infected individuals during the first and second waves of COVID-19 reasonably well~\cite{gopal,gopal2}. To predict and to identify the evolution of COVID-19, a series of research has recently been carried out~\cite{lin,li,he,gopal,gopal2,post,kavi,chen,ranjan,savi,nld}. In particular, the SIR and linear fractal-based models have predicted the daily active cases of the outbreaks of COVID-19 in India~\cite{post,kavi}. Additionally, the dynamic evolution of the SEIR model along with various parameters, like incidence rate, transmission rate, test positivity rate, case fatality rate and intervention parameters,
were demonstrated to examine the first and second waves of COVID- 19~\cite{chen,ranjan,mus}. Also, the SEIR model was modified by Suwardi et al. ~\cite{su} to investigate the spreading pattern of COVID-19 by considering vaccination and isolation as critical parameters. However, recent studies have also shown that the effectiveness of vaccines is getting reduced over time~\cite{waning,wan2,wan3}.

However, to eradicate  COVID-19 so that it becomes effectively endemic, the government has taken various health and social measures such as surveillance, contact tracing, isolation of infected individuals, and insistence of protective behaviors to fight COVID-19. In addition, the COVID-19 vaccines are treated as essential tools to stop the global spread of the pandemic. In general, the COVID-19 vaccines have been found to be very  effective in preventing severe illness, hospitalization and death from almost all the  current virus variants. The effectiveness of the vaccines has been found to depend upon many parameters such as the possible emergence of other variants and the number of vaccinated individuals. But, still, some threats occur both individually and at the societal level as there is a chance of people getting infected even after completing the vaccination due to the possibility of vaccine waning among the people after a certain period. Therefore, studies related to the rate of vaccination or vaccine weakening along with the SEIR model are required to be investigated on the  importance and the need for additional booster vaccinations.

Therefore, in this present study, we aim to develop a modified SEIR model that includes additional compartmental for vaccinated individuals, to investigate the number of infected individuals with the inclusion of the rate of vaccination and vaccine waning capacity in the public health system in order to support control measures.

The main aim of this paper is to interpret the underlying general dynamics of our proposed model with vaccines allowed to wane, and then check the consistency of the model by studying the dynamics of COVID-19 with  normalized populations. Further, our study is extended to examine the evolution of the number of infected individuals in India. In particular, we investigate the role of vaccines in controlling the disease spread and how they will influence the future. We also examine the previous waves of infection due to COVID-19 with our model and whether it is consistent with real-time data. We use these results to predict the possible waves of COVID -19 in the future. Furthermore, the variability due to the rate of vaccination has also been studied. The transmission rate is also taken to be a time-varying one, and it depends both on governmental actions such as travel restrictions and lock-downs ~\cite{gopal, gopal2,lin,li,savi} as well as on the public perception of risk. Finally, our study also investigates achieving herd immunity by vaccinated individuals. It indicates that the role of vaccines in attaining herd immunity is also quite crucial to our results. With proper booster vaccination campaigns, the vaccine waning could  be overcome, and the endemic state with herd immunity may become feasible.

With the above-mentioned objectives,  Section II describes our proposed description of the SIERV model and its dynamical analysis. We present the number of infected individuals with COVID-19 in India under the variability of vaccination and its associated parameters in section III. In Section IV, we present an analysis of herd immunity to COVID-19 in India. Finally, we summarize our findings in Section V.

\begin{figure}
	\centering
	\includegraphics[width=1.0\columnwidth]{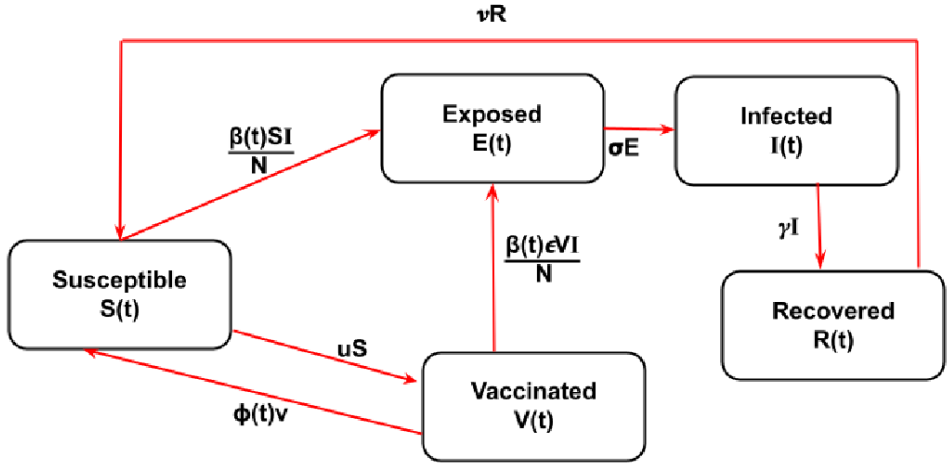}
	\caption{Block diagram depiction of interaction of various population compartments based on SIERV model}
\label{fig1}
\end{figure}

\section{Analysis of the SIERV model}  
\label{Dom}
To investigate the importance of natural immunity, vaccine waning, and the rate of vaccination, we modify and extend the SEIR model~\cite{lin,li,gopal,gopal2,he,savi}, by considering the additional compartment of vaccinated group $V(t)$, which denotes the number of vaccinated individuals. We call the modified model as the SEIRV model.  Therefore, the suggested  SEIRV  model is governed by the following set of coupled first order nonlinear ordinary differential equations(ODEs),\begin{subequations}

\begin{eqnarray}
&&\dot{S}=-u{S}-\beta(t)\frac{SI}{N}+\phi(t) {V}+\nu {R},  \\
\label{eq1b}
&&\dot{E}=\beta(t)\frac{SI}{N}-\sigma E +\epsilon \beta(t)\frac{VI}{N},  \\
\label{eq1c}
&&\dot{I}=\sigma E-\gamma I   , \\
&&\dot{R}=\gamma I-\nu{R},  \\
&&\dot{V}=u{S}-\phi(t) {V} -\epsilon\-\beta(t)\frac{VI}{N} \\
&&\dot{D}=\gamma d I-\lambda D ,  \\
&&\dot{C}=\sigma E.  
\label{eq1}
\end{eqnarray}
%\label{eq1}
\end{subequations}

\begin{figure}
	\centering
	\includegraphics[width=1.0\columnwidth]{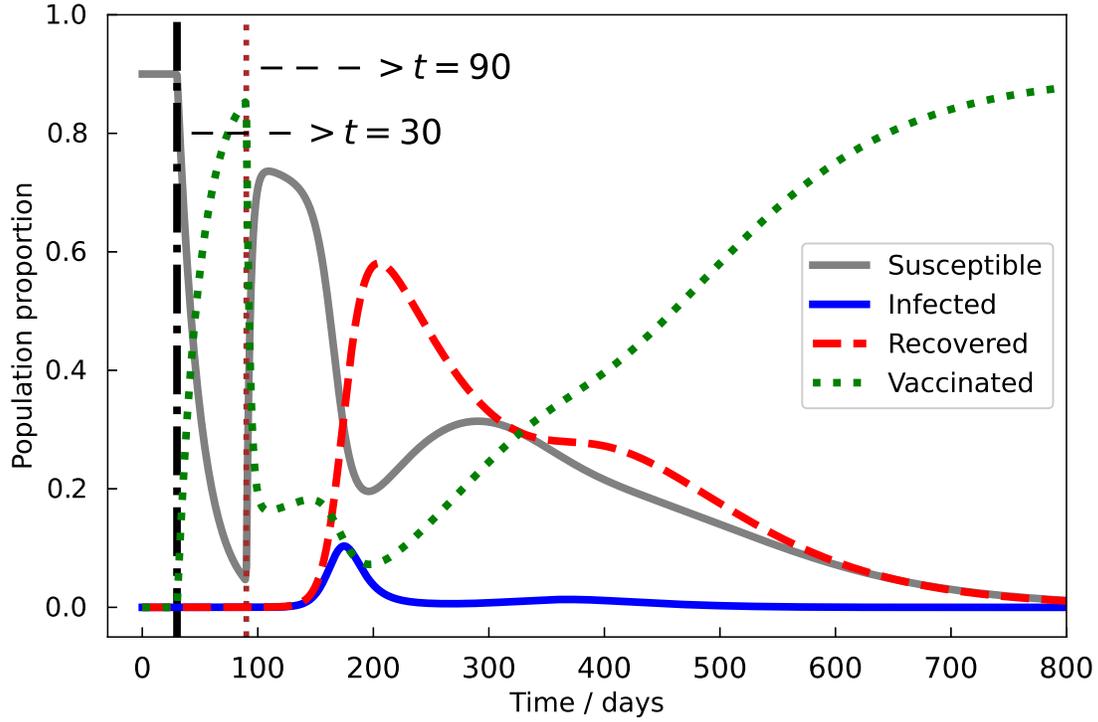}
	\caption{Numerical simulation of our SIERV model describes proportion of the susceptible $S(t)$, infected individuals $I(t)$, recovered $R(t)$ and vaccinated $V(t)$  population with parameters given in table \ref{table2}.}
\label{fig2}
\end{figure}

 In Eq.(\ref{eq1}), $S$, $E$, and $I$ represent the susceptible population, exposed population and currently infected population (excluding the recovered and dead cases), respectively. Also, $R$ denotes the population removed, including recovered and death numbers, and $V$ is the number of vaccinated people. Further, $N$ is the total number of the population, which includes two more categories: $D$ is a public perception of risk concerning severe cases and deaths, and $C$ is the number of cumulative cases (including both reported and unreported). Based on this, the abstract model of the COVID -19 dynamics is presented in Fig. \ref{fig1}, representing the above set of coupled equations (\ref{eq1}). Further, the following parameters are also considered for the dynamical equations (\ref{eq1}) : $u$ denotes the rate of vaccination, that is, the total number of people vaccinated per day, and $\nu$ is the term governing the loss of natural immunity, $\phi(t)$ is the term that determines vaccine waning, and $\gamma$ is the mean infectious period. Also, $\sigma$ is the mean latent period, while $d$ denotes the proportion of severe cases and $\lambda$ is the duration of public reaction.

Then, the coefficient $\beta(t)$  is considered to be a time-varying disease transmission rate  ~\cite{lin,he,savi,gopal} which incorporates  the impact of governmental action (1 - $\alpha$) and the individual action is denoted by $(1-D/N )^k$. Here, $\alpha$ and $k$ denote governmental action strength and intensity of individual action strength, respectively. These values are adjusted in each of the specific lock-down periods in 2020 and unlock periods. Then the transmission rate which is defined as~\cite{lin}
\begin{align}
\beta(t)=\beta_{0}(1-\alpha) \left (1-\frac{D}{N}\right)^{k}. 
\label{eq3}
\end{align} 
Moreover, in Eq.(\ref{eq1}), the vaccine waning  term $\phi(t)$ depends on various parameters such as vaccine availability, immune response and food habitat of people. The short development history and lack of long-term follow-up studies  for vaccine waning in the mathematical model makes it less likely to predict clearly the future evolution of COVID-19 among the people. However, to investigate the effect of the vaccine, and potential effects of behavioral response, the term $\phi(t)$ is considered through  parameterization of generalized logistic function. The variation of $\phi(t)$ with time can be represented as
\begin{align}
\phi(t)= \frac{\phi_{0}L}{1+e^{-m_0(t-t_0)}},
\label{eq2w}
\end{align}
where $\phi_{0}$ is the mean vaccine waning, $L$ is the initial protection provided by the vaccine, $m_0$ is the initial rate of vaccine waning and $t_0$ is the time at which the vaccine starts to wane since being vaccinated. The value of $\phi(t)$ in (\ref{eq2w}), varies from $\phi_{0}\frac{L}{2}$ at $t=t_{0}$ to $\phi_{0} L$ as $t\rightarrow \infty$.

 To start with, we have presented the initial numerical results of our mathematical model with the parameters as mentioned in the table \ref{table2}. Our results show that our proposed epidemic model (\ref{eq1}) attains an endemic equilibrium, once enough population gets affected and immunized against the disease (See Fig. \ref{fig2}).  For instance, once the vaccination starts at $t=30$, there is a dip in susceptibility, as the people get immunized through vaccination. In continuation, the vaccine begins to wane when $t=90$. There is an increase in disease exposure (more number of people susceptible), which makes the disease to spread quicker and causes a peak of infection. After a subsequent time, the number of susceptible people starts to drop and approaches 0 due to the continuous vaccination  (See Fig. \ref{fig2}).

\begin{figure*}
\centering
\includegraphics[width=1.0\columnwidth]{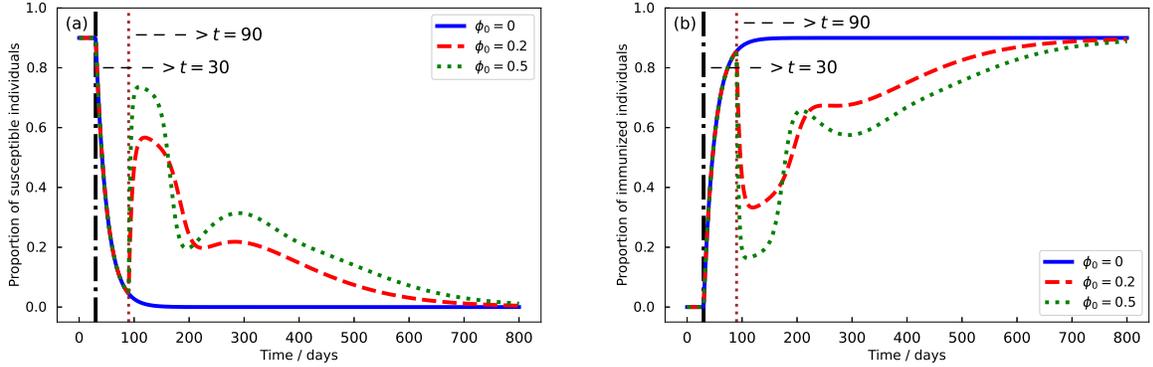}
\caption{Numerical simulation of the proportion of number of the (a) susceptible individuals $S(t)$, and (b) immunized (including both vaccinated and recovered individuals, $V(t)+R(t)$) individuals for different values of mean vaccine waning $\phi_{0}$. The other parameters are mentioned in table in \ref{table2}.}
\label{fig4}
\end{figure*}

 In Fig. \ref{fig4}, we have studied the dynamics of the number the susceptible and vaccinated people with respect to the different values of the mean waning of the vaccine ($\phi_0$) with help of Eqs.(\ref{eq1})-(\ref{eq2w}). When we have a vaccine which is not subjected to waning, i.e, $\phi_0=0$, the number of susceptible people drops down and approaches zero and a potential epidemic can be prevented and this is denoted by the solid line in Fig. \ref{fig4}(a). Once the vaccine starts to wane among the population (for instance,  $\phi_0 = 0.2$ and  $0.5$), there will be an increase in the susceptibility (see Fig.(\ref{fig4})(a)) and it will create a possibility of subsequent waves of infection rather than a new variant. Fig. \ref{fig4}(b) shows a scenario in which the number of vaccinated individuals for different values of vaccine waning, namely $\phi_0 = 0, 0.2$ and  $0.5$. From this one can observe that there is an initial dip in the number of vaccinated individuals due to vaccine waning ($\phi_0= 0.2$ and  $0.5$), but due to the continuous and efficient vaccination over time and absence of waning ($\phi_{0}=0$), the number of vaccinated individuals approaches a stable steady state of maximum value close to 1.

The scenarios illustrated in Figs. \ref{fig4}(a) and \ref{fig4}(b) confirm an endemic steady-state, which corresponds to a fraction of the population of susceptible and vaccinated ones. Here, the steady-state of endemic equilibrium illustrates an essential measure in the COVID -19 analysis and is denoted as a basic reproduction number that is represented as $\Re_{0}$.  We calculate $\Re_{0}$ by using the next generation matrix concept, where initially we the obtain the steady state vector with disease-free endemic state when $I=0$ in Eq.(\ref{eq1}), 
\begin{align}
 X_{0}=(\frac{\phi_0 V_0}{u} , 0,0,0, V_0 ,0,0 ),
\label{eq4}
\end{align} 
From this steady state vector, we can estimate the basic reproduction number $\Re_{0}$ by using the next generation matrix concept~\cite{nxt}, since $\Re_{0}$ measures as the expected number of secondary infections produced by an index case (an initial infected cases) and depends on the completely exposed population by typical infected individuals~\cite{van,diek}. Therefore, we consider $X = (E,I)^T$, then one can write Eq.(\ref{eq1}) as,
 \begin{align}
 \dot {X}= \mathbb{F}(X) - \mathbb{W}(X),
\label{eqM}
\end{align}
where $ \mathbb{F}(X)=(\beta S_0I/N + \epsilon \beta V_0I/N , 0 )^T$ and $ \mathbb{W}(X)=(-\sigma E, \sigma E - \gamma I)^T$ and the corresponding jacobian for the disease free steady state takes the form, 
\begin{align}
\ F=\left( \begin{array}{cc} 0 & \frac{\beta(\epsilon V_0 +S_0)}{N} \\
0 & 0 \end{array} \right),
\label{eqM}
\end{align}
\begin{align}
\ W^{-1}=\left( \begin{array}{cc} \frac{-1}{\sigma} & 0 \\
\frac{-1}{\gamma} & \frac{-1}{\gamma} \end{array} \right),
\label{eqM2}
\end{align} 
Then, 
\begin{align}
\ FW^{-1}=\left( \begin{array}{cc} -\frac{\beta(\epsilon V_0 +S_0)}{N\gamma} & -\frac{\beta(\epsilon V_0 +S_0)}{N\gamma} \\
\ 0 & 0 \end{array} \right),
\label{eqM2}
\end{align} 
Here, the dominant eigen values of the matrix $FW^{-1}$ gives the value of $\Re_{0}$ which becomes, 
\begin{align}
\Re_{0}={ \frac{\beta(S_0+\epsilon V_0)}{N\gamma}},
\label{R0_1}
\end{align}
Substituting the value of $S_0$ from Eq.(\ref{eq4}) yields the following expression 
\begin{align}
\Re_{0}={\frac{\beta V_0 (\phi_0+\epsilon u)}{u N\gamma}},
\label{R0}
\end{align}
and this form helps to identify the disease transmission  and also tells us how many secondary infections that a disease can generate.  For instance, if  $\Re_0>1$ then the transmission will be exponential; otherwise, the disease will go extinct if $\Re_0 < 1$~\cite{r1,r2}.

Further, disease extinction or  persistence can also be determined by using the stability of the endemic equilibrium of the model (\ref{eq1}). Let us assume the disease free $X_0$ is asymptotically stable in Eq.(\ref{eq1}) by considering  $V=N-S-I-E-R$. So, we can eliminate $V$ from the system,  and the corresponding Jacobian matrix becomes
\begin{align}
\ J(X_0)=\left(\begin{array}{rrrr}
-u & 0 & -K_{1} & \nu \\
0 & -\sigma & (K_{1} + K_{2}) & 0 \\
0 & \sigma & -\gamma  & 0 \\
0 & 0 & \gamma & -\nu %\nonumber
\end{array}\right),
\end{align}
where $K_1=\beta S_0/N$ and $K_2=\beta \epsilon V_0/N$. The matrix $J(X_0)$ has four eigenvalues and they are 
\begin{align}
\lambda_1&=-\nu,\nonumber \\
\lambda_2&=-u,\nonumber \\
\lambda_3&=\frac{-1}{2}(\frac{\beta}{\Re_{0}}-\sigma+\sqrt{(\frac{\beta}{\Re_{0}})^{2} -2{\left(\frac{\beta}{\Re_{0}}- 2\,(K_{1}-\,K_{2})\right)} \sigma + \sigma^{2}}),\nonumber\\
\lambda_4&=\frac{-1}{2}(\frac{\beta}{\Re_{0}} -\sigma -\sqrt{(\frac{\beta}{\Re_{0}})^2 -2{\left(\frac{\beta}{\Re_{0}}- 2\,(K_{1}-\,K_{2})\right)} \sigma + \sigma^{2}}).\nonumber
\end{align}
%\end{subequations}
%%%%%%%%%%%%%%%%%%%%%%%%%%%%%%%%%%%%%%%%%%%%

When $\Re_{0}<1$, all the eigenvalues are negative and consequently, $X_0$ is locally asymptotically stable.

\begin{table}
\centering
\caption{Model parameters chosen for the general analysis of Eq.~(\ref{eq1}) }
\label{table2}
\begin{tabular}{lll}
\hline\noalign{\smallskip}
%\toprule
Parameter & Description & value/remarks/reference \\
%\toprule 
\noalign{\smallskip}\hline\noalign{\smallskip}

$N_{0}$ & Initial number of population & 1(Normalized constant)\\

$S_{0}$ & Initial number of susceptible population & 0.9$N_0$ (constant)  \\

$E_{0}$ & Exposed persons for each infected person & $10I_{0}$\ \\
 
$I_{0}$ & Initial state of infected persons  & $1 \times 10^{-6}$ \\
 
$\alpha$ & Government action strength  & 0\\

k & intensity of individual reaction & 0\\

$\sigma^{-1}$ & Mean latent period & 3 (days) \\
 
$\gamma^{-1}$ & Mean infectious period & 5 (days) \\ 

$d$ & Proportion of severe cases & 0.2  \\

u & Vaccination rate & 0.05\\ 

$\epsilon$ & Vaccine inefficacy & 0.3 \\ 
 
$L$ & Maximum protection vaccine provides initially &  1 \\

$\phi_0$ & Mean vaccine waning &  0.5 \\

$t_0$ & Start of vaccine waning &  60 days from being vaccinated \\

$\lambda^{-1}$ & Mean duration of public reaction & 11.2 (days)  \\

%\toprule
\noalign{\smallskip}\hline
\end{tabular}
\end{table}

\begin{figure*}
	\centering
	\includegraphics[width=1.0\textwidth]{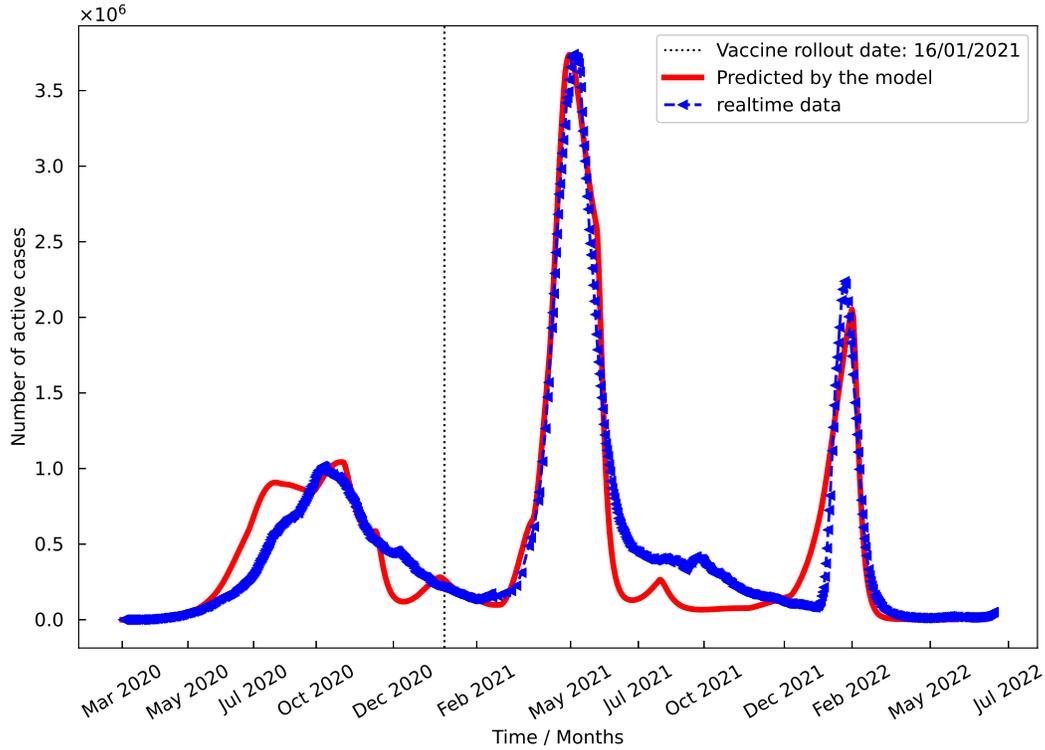}
	\caption{Numerical simulation of the number of infected individuals (after removing the number of recovered people on a particular day). The curves represent the numerical simulation of the number of infected individuals (active cases) from Feb 1, 2020 to Apr 30,  2022, using the SEIRV mathematical model. Data available between Feb. 2020 to Apr. 2021 are taken for fitting the parameters. The red curve indicates the real number of infected individuals (active cases) and the vertical black dotted line indicates the start of vaccination campaign in India. The other parameters and initial conditions are mentioned in table \ref{table1}.}
\label{fig5}
\end{figure*}

\begin{table}
\centering
\caption{Model parameters for studying COVID-19 spread in India as per Eq.~(\ref{eq1})}
\label{table1}
\begin{tabular}{lll}
\hline\noalign{\smallskip}
%\toprule
Parameter & Description & value/remarks/reference \\
%\toprule 
\noalign{\smallskip}\hline\noalign{\smallskip}
$N_{0}$ & Initial number of population & India~\cite{gov1}\\

$S_{0}$ & Initial number of susceptible population & $0.9N_{0}$ (constant)  \\

$E_{0}$ & Exposed persons for each infected person & $20I_{0}$\cite{gov1} \\
 
$I_{0}$ & Initial state of infected persons & 3 (India)~\cite{gov1} \\
 
$\alpha$ & Government action strength  & varied in lock-down/unlock period\\

k & intensity of individual reaction & 1117.3~\cite{lin,savi} \\

$\sigma^{-1}$ & Mean latent period & 3 (days) \\
 
$\gamma^{-1}$ & Mean infectious period & 5 (days) \\ 

$\gamma_{R}^{-1}$ & Delayed removed period & 22 (days) \\

$d$ & Proportion of severe cases & 0.2  \\

u & Vaccination rate & 2.7 million vaccinations per day \\ 

$\epsilon$ & Vaccine inefficacy & 0.1 \\ 
 
$L$ & Maximum protection vaccine provides initially &  1 \\
 
$ \phi_0 $ & Mean vaccine waning & $9 \times 10^{-4}$ \\

$t_0$ & Start of vaccine waning &  100 days from being vaccinated \\

$\lambda^{-1}$ & Mean duration of public reaction & 11.2 (days)  \\

%\toprule
\noalign{\smallskip}\hline
\end{tabular}
\end{table}

%\section{Results and Discussion}

\label{Rnd}
\section{COVID-19 Analysis in India}

\subsection{Dataset}
All the data sets used in this manuscript are publicly available. We have used the publicly available COVID-19 Data Repository from the worldometer website \\ \underline{(https://www.worldometers.info/coronavirus/country/india/)}. \\
Further, we have taken the daily number of infected individuals with COVID-19 and the number of vaccinated individuals from the official COVID-19 data Website managed by the Ministry of Health and Family Welfare (MoHFW), Government of India \underline{(https://www.mohfw.gov.in/)}. Using these data, we have analyzed our SEIRV model further in the following.

\subsection{Analysis of real data:}

To justify  our results and to explore additional important properties of the model, we fitted the model  to real time COVID-19 data of India based on the parameters mentioned in Table \ref{table1} and then we carried out a numerical simulation. Initially, we started with the real-time daily number of infected individuals (active cases) of COVID-19 in India ~\cite{gov1} and fitted the data with our model (\ref{eq1}) with the help of appropriate governmental and individual action parameters. 
 
 \par Figure \ref{fig5} shows the evolution of the number of infected individuals since the onset of COVID-19 in India. It clearly shows that after the second wave of COVID-19, there has been a sudden drop in the infections due to governmental actions such as lockdowns, travel restrictions, and vaccination. One can also state that the severity of the disease and the death rate dropped after the vaccination campaign. However,  there was also a surge in infection from Feb 2022 - to March 2022. It shows that the infection rates were significant mainly due to public risk perception and relaxation in the  restrictions imposed by the government. However, one can clearly surmise that the severity of the infections, which resulted in lowered hospitalizations and a low-value of mortality rate during this period, is mainly due to the effect of vaccination.

Further, we have also observed that the severity of infection was again lowered up to half a time of the second wave (See Fig. \ref{fig4}). However, the government has  almost removed all the restrictions. It indicates that it strongly advocates the  success of vaccine in holding down the infection rates. 

From the above analysis, we need to examine the performance of our mathematical model (\ref{eq1}) and its prediction accuracy. Since the data-set for COVID-19 is growing every day, the difference between the actual and fitted data also has enormous magnitudes. Therefore, we used the statistical measure, namely the normalized root mean squared value error (ReMSE)~\cite{yang}, and it is defined as
\begin{align}
R e M S E=\sqrt{\frac{{\frac{1}{T} \int_{0}^{T} (I_{a}(t)-I_{e}(t))^{2}  dt}}{{\max (I_{a})}}}
\end{align}
Here $I_{a}$ is the real-time data of the total number of infected individuals, and the $I_{e}$ is the number of infected individuals predicted by the proposed model. From this, the value of ReMSE is found as 0.0442, and in percentage, it is found to be 4.42\% confirming the reliability of the model (\ref{eq1}).

\begin{figure*}
	\centering
	\includegraphics[width=1.0\textwidth]{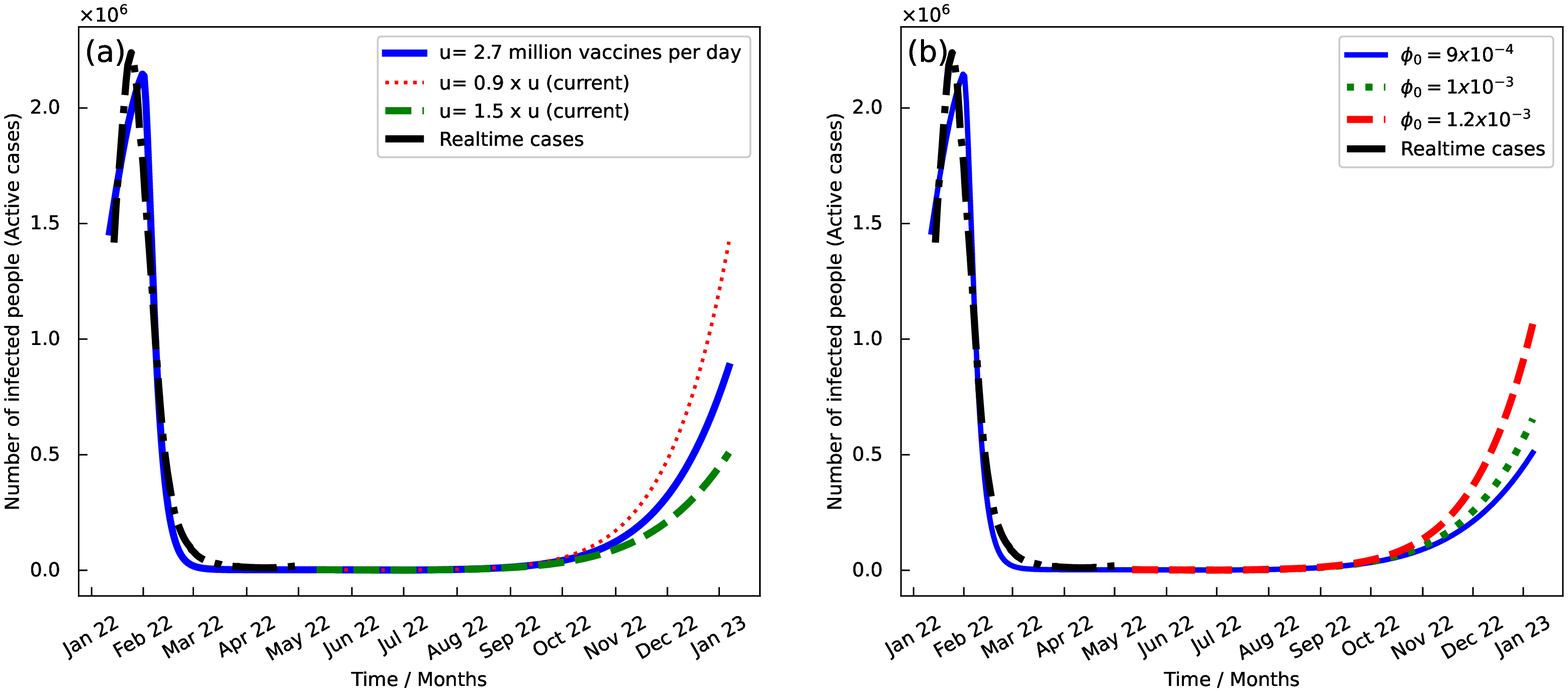}
	\caption{(a) Number of infected individuals from Jan 2022 to Jan 2023 with various vaccination rates. Here, the black-dash dotted curve indicates the real-time data taken from website~\cite{gov1,gov1}, and the solid line indicates the simulation carried out for the current vaccination rate. The dotted (red) and dashed (green) curves are for low and high values of the vaccination rate. (b) The number of infected individuals from Jan 2022 to Feb 2023 concerning the variation of vaccine waning. Here black-dash dotted curve indicates the real-time date taken from website~\cite{gov1}, the solid blue line low vaccine waning $\phi_0=9$x $10^{-4}$. The dotted and dashed curves are denoted  for $\phi_0= 1$x $10^{-3}$ and $\phi_0= 1.2$ x $10^{-3}$, respectively.}
\label{fig6}
\end{figure*}

\subsection{Effects of the vaccination rate and vaccine waning in India}

In India, the average growth rate of total COVID -19 cases was high during the second wave of COVID -19 when compared with the first and third waves. The average growth rate of active cases in the individual states was highest in Maharashtra and lowest in Nagaland. Similarly, the average growth rate of hospitalization was highest in both the first and second waves of COVID -19, and it got significantly lowered in the third wave of COVID -19. This is essentially due to effect of vaccination, and it has significantly slowed down the growth rate of COVID -19 cases and hospitalization. Therefore, we have planned to investigate the impact of vaccination and vaccine waning rates in the model (\ref{eq1}).

As of the first week of May 2022, around 862 million people were fully vaccinated in India, since the vaccination campaign started on 16$^{th}$ January, 2021. The total vaccinations  account for up to 62.5\% of the total population (May 2022), and the vaccination is being done at approximately 2.7 million per day approximately~\cite{gov1}. This value is taken as the vaccination rate $u$ in our study in Eq.(\ref{eq1}). Still, there are 37.5\% of people left to be fully vaccinated as on May 04, 2022. These unvaccinated people are still susceptible to infections. In the present study, we have analyzed the number of infected individuals for different vaccination rates. Fig. \ref{fig6}(a) shows the number of infected individuals over time for different vaccination rates. 

Initially, the occurrence of the number of the individuals is estimated with the current value of the vaccination rate $u$ with various model parameters given in table \ref{table1}. Our results on the number of infected individuals match with actual data, and we note that there will be an increase in the new infection around December 2022 (See Fig. \ref{fig6}(a)). By enhancing the vaccination rate by 50\% (that is to 1.5u), an effective decrease occurs in the infected individuals, and new infections will be delayed. Furthermore, by reducing the vaccination rate of $u$ by 10\% (to $0.9u$), a significant increase in the number of infected individuals will occur earlier in October 2022 (See Fig. \ref{fig6}(a)).
Biologically, vaccines contain anti-virus and they are responsible for boosting the immune system of suspects, which will raise the recovery rate, resulting in a drop in the infected population. However, the occurrence of new peaks in the infected individuals is due to the vaccine-waning effect and public perception of risk. Further, one can subsequently reduce the vaccine waning effects by booster vaccination doses. This will avoid the chance of future peaks. It also decreases the probability of the virus mutating into new variants because viruses of infectious diseases like COVID-19 are prone to mutations~\cite{mut} as long as they keep making new infections.
 Based on the estimates of recent study~\cite{waning,wan2,wan3}, the defensive antibodies produced in our body due to the vaccine have been steadily waning over time, resulting in a probability of creating new infections. Further, studies suggest that the defensive antibodies were reduced to approximately one-half of the initial state after five months from the date of vaccination~\cite{waning,wan2,wan3}. It causes more people to be vulnerable to infection. It will result in huge costs individually and for society because the more the people become susceptible, the more the infection will occur, resulting in an epidemic scenario. However, there is uncertainty about how fast the vaccine wanes after an year. Figure \ref{fig6}(b) demonstrates the number of infected individuals versus time by considering mean vaccine waning values. For instance, the real data of the number of infected individuals match with our model (\ref{eq1}) results for $\phi_0= 9 \times 10^{-4}$, and it is indicated as a solid curve. On increasing the values of the mean vaccine waning to $\phi_0= 1 \times 10^{-3}$ and $\phi_0= 1.2 \times 10^{-3}$, one observes the corresponding significant increase in the number of infected individuals (See dotted and dashed lines in Fig.\ref{fig6}(b). Therefore one may note that the increased vaccine waning will result in an increase in the number of infections of COVID-19.

\begin{figure}
	\centering
	\includegraphics[width=1.1\columnwidth]{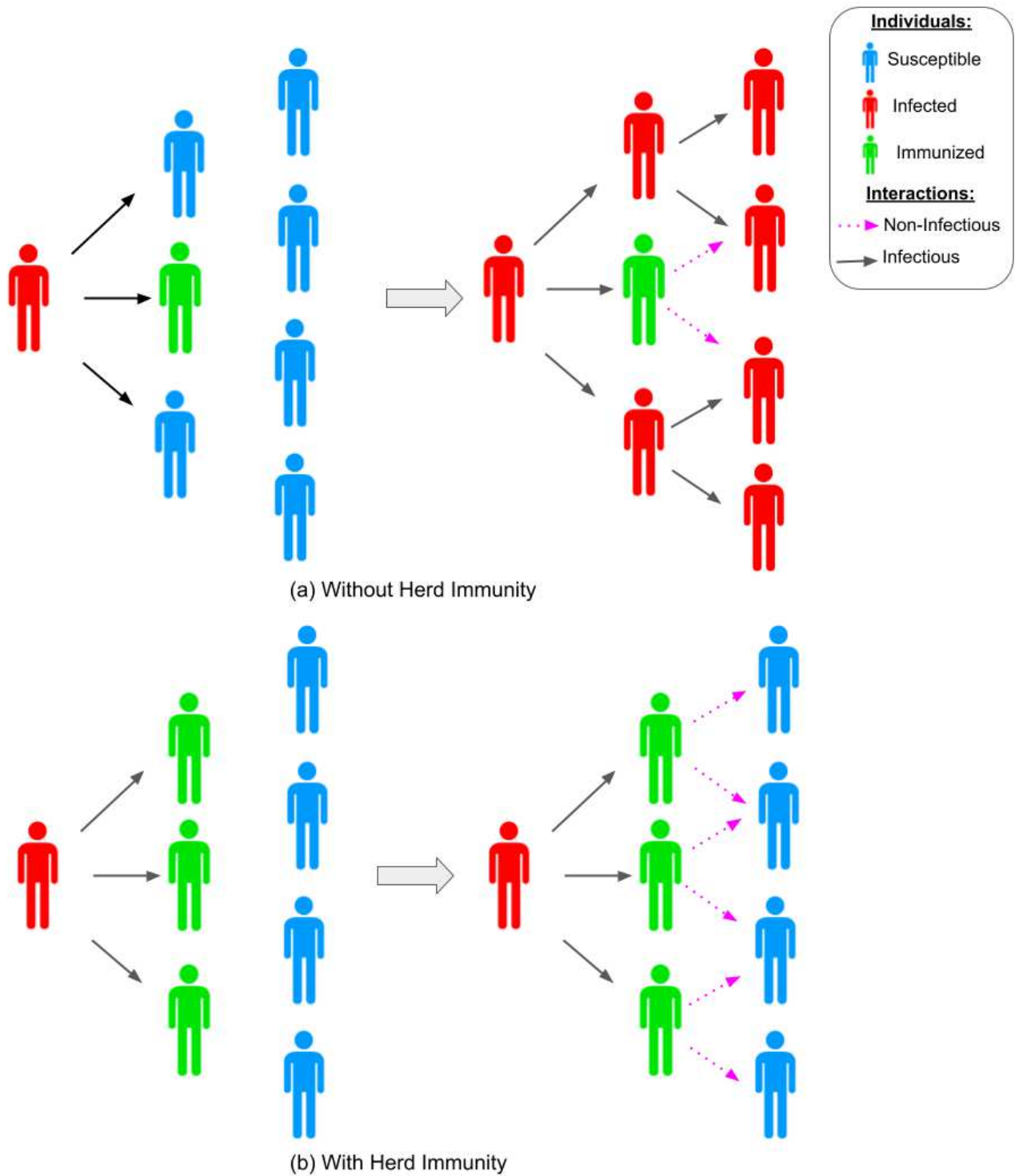}
	\caption{Schematic representation of the people without and with herd immunity: (a) without herd immunity: In this case, one can see that the primary (vertical) layer thoroughly infects the secondary (vertical) layer, (b) with herd immunity: the secondary (vertical) layer is offers indirect protection by the completely immunized primary layer, which breaks the chain of disease transmission.}
\label{herd}
\end{figure} 
 
\section{Analysis of possible herd immunity for COVID -19 in India}

Herd immunity is associated with a scenario in which people develop immunity against a contagious infectious disease that manifests when they are immune, either through vaccination or due to previous infection, and become resistant to that disease. This provides some amount of indirect protection for those who are not immune to the disease ~\cite{herdi,kwok,gowri}. Fig. \ref{herd} depicts the representation of people without and with herd immunity. However, to achieve herd immunity against COVID-19, a substantial proportion of the population needs to be vaccinated. Also, after a certain intense period, herd immunity can be developed through the natural course of infection~\cite{gowri}.

Further, the concept of herd immunity expects that infectious diseases can be controlled or eradicated once the total immunized population reaches the herd immunity threshold level. Recent studies also point out that herd immunity~\cite{herdi,strogatz}  is achieved when one infected person in a population generates less than one secondary case on an average. It represents the situation where the adequate reproduction number $\Re_{0}$ dropping below one without interventions.

\begin{figure*}
	\centering
	\includegraphics[width=1.0\textwidth]{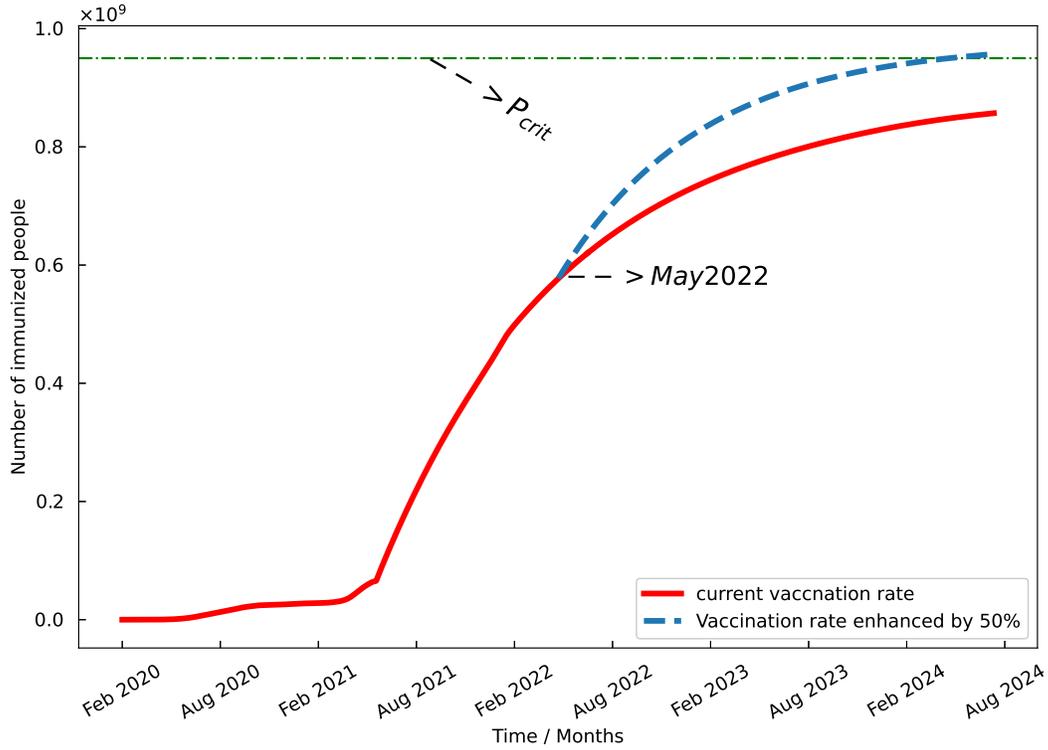}
	\caption{The number of immunized people (vaccinated and recovered population combined) with respect to time in months. Here, $P_{crit}$ denotes the critical population for herd immunity. In the figure, red solid line indicates the evolution of immunized people into the future with the current vaccination rate, and the blue dashed line indicates the future evolution with 1.5 times the current vaccination rate.}
\label{fig7}
\end{figure*}

Therefore, if we designate the term $\ P_{crit}$  as the total number of people who are immune to disease either by vaccination or by naturally getting affected by the virus, it may be represented as, 

\begin{align}
\ P_{crit}=\frac{1}{1-\epsilon} \left (1-\frac{1}{\Re_0}\right),
\label{eq9}
\end{align} 
where ($1 - \epsilon$) is the vaccine efficacy rate and $\Re_{0}$ is the effective reproduction number.
 As in the cases of general vaccines, in most of the vaccines of COVID-19, the vaccine efficacy is around 0.9~\cite{effi}. 
As of now, 62.5\% of Indians are fully vaccinated. For SARS-CoV-2, most of the estimates of $\Re_0$ of India are in the range 2.5–4, with no explicit agreement as $\Re_0$ varies during the time of events where mass gatherings were held. So, we fixed $\Re_0=2.8$, and the herd immunity threshold for SARS-CoV-2 is expected to require 69\% for population immunity. This is denoted by the dashed line in Fig.(\ref{fig7}). As the vaccine waning is considered, even if a person has completed two doses of the vaccine, the vaccine will start to wane. 

 Further, we also present our results of the total number of immunized people, including the recovered and vaccinated ones, with the current vaccination rate since the vaccination campaigns started. From this, one can infer that there is a steady increase in the number of immunized people and it does not approach the $P_{crit}$ state (See the solid line in Fig. \ref{fig7}). However, when the vaccination rate is enhanced by 50\%, and the number of immunized people intersects the $\ P_{crit}$ value represented by a dashed line, it tells that a complete endemic state will be reached through herd immunity around early 2024 using the faster vaccination rate (See the dashed line in Fig. \ref{fig7}). Therefore, our results strongly advocate that speeding up the vaccination rate is a major route to achieving an endemic state.

\section{Conclusion}
We have proposed an epidemic model for pandemic COVID-19 in India that represents the susceptible population, exposed population, infected population, and recovered population and vaccinated population. First, we study the proposed model's dynamic properties and obtain a stable endemic equilibrium.  This value is stable if the primary reproduction number is less than unity which means that the disease is eradicated asymptotically from the population. Further, our results also summarize that our proposed SIERV model serves as an effective tool to study the future of pandemics when the vaccine we have at hand wanes over time. Our studies with various scenarios of vaccination rates and estimating the critical population for achieving herd immunity show that this epidemic is quite far from over. Moreover, the individual's perception of risk is also important because it can result in a higher infection rate without adhering to the usual COVID-19 protocols such as masking, social distancing and avoiding mass gatherings. Furthermore, an estimation of the herd immunity shows that if the vaccination rate were increased and more people became vaccinated than the threshold of the critical population, then the infection reproduction rate would decrease. This means that the chain reaction of infections gets controlled, and an epidemic situation can be avoided. The results also prove that unless or until we have a  vaccine that does not wane over time, the infection rate may get higher, and people will get infected even if vaccinated. Finally, our  results show that the booster vaccination campaigns  may be effective in protecting people from infection. 
\\ 

{\bf Data Availability Statement:}The data sets on the current study are available from the corresponding author on reasonable request

\section*{Acknowledgements}
The work of V.K.C. and R.G. forms part of a research project sponsored by SERB-DST-CRG Project  Grant No. C.R.G./2020/004353. R.G. and V.K.C. thanks DST, New Delhi, for computational facilities under the DST-FIST programme (Grant No. SR/FST/PS-1/2020/135) to the Department of Physics. M.L. wishes to thank the Department of Science and Technology for the award of a DST-SERB National Science Chair (N.S.C./2020/000029).

\end{document}